\documentclass[aps,prc,amsmath,showpacs,floats,twocolumn,superscriptaddress,byrevtex]{revtex4}

\usepackage{graphicx,epsfig}

\newcommand{\beqy}{\begin{eqnarray}}
\newcommand{\eeqy}{\end{eqnarray}}
\newcommand{\bmlet}{\begin{subequations}}
\newcommand{\emlet}{\end{subequations}}

\begin{document}
\title{Further explorations of Skyrme-Hartree-Fock-Bogoliubov mass formulas.\\
II: Role of the effective mass.}
\author{S. Goriely}
\affiliation{Institut d'Astronomie et d'Astrophysique, 
Universit\'e Libre de Bruxelles - CP226,
1050 Brussels, Belgium}
\author{M. Samyn}
 \affiliation{Institut d'Astronomie et d'Astrophysique, 
Universit\'e Libre de Bruxelles - CP226,
1050 Brussels, Belgium}
\author{M. Bender} 
\affiliation{Service de Physique Nucl\'eaire Th\'eorique et de Physique 
Math\'ematique, Universit\'e Libre de Bruxelles 
- CP229, 1050 Brussels, Belgium}
\author{J. M. Pearson}
\affiliation{D\'ept.\ de Physique, Universit\'e de Montr\'eal, 
Montr\'eal (Qu\'ebec), H3C 3J7 Canada}

\begin{abstract}
We have constructed four new complete mass tables, referred to as
HFB-4 to HFB-7, each one including
all the 9200 nuclei lying between the two drip lines over the range of $Z$ and
$N \ge 8$ and $Z \le 120$. HFB-4 and HFB-5 have the isoscalar effective mass
$M_s^*$ constrained to the value $0.92M$, with the former having a 
density-independent pairing, and the 
latter a density-dependent pairing. HFB-6 and HFB-7 are similar, except that
$M_s^*$ is constrained to 0.8$M$. The rms errors of the mass-data fits are
0.680, 0.675, 0.686, and 0.676 MeV, respectively, almost as good as for the 
HFB-2 mass formula, for which $M_s^*$ was unconstrained.
However, as usual, the single-particle spectra 
depend significantly on $M_s^*$. This
decoupling of the mass fits from the fits to the single-particle spectra has
been achieved only by making the cutoff parameter of the $\delta$-function
pairing force a free parameter. An improved treatment
of the center-of-mass correction was adopted, but although this makes a
difference to individual nuclei it does not reduce the overall rms error
of the fit. The extrapolations of all four new mass formulas out to the
drip lines are essentially the same as for the original HFB-2 mass formula. 
\end{abstract}

%\begin{keyword}
%NUCLEAR STRUCTURE: Binding energies and masses, Hartree-Fock
% PACS Numbers: 21.10.Dr Binding energies and masses , 
%               21.30.-x Nuclear forces, 
%               21.60.Jz (HF & QRPA)
%\end{keyword}


\maketitle
\newpage

%******************************************
\section{Introduction}
%*****************************************

In the last few years it has become possible to construct complete mass tables 
by the Hartree-Fock (HF) method \cite{sg01,ms02,sg02,ms03}, with the 
parameters of the underlying forces being fitted to essentially all of
the available mass data.
These HF calculations are based on conventional Skyrme forces of the form
\begin{eqnarray}
\label{1}
v_{ij} 
& = & t_0(1+x_0P_\sigma)\delta({{\bf r}_{ij}}) \nonumber\\
& &+t_1(1+x_1P_\sigma)\frac{1}{2\hbar^2}\{p_{ij}^2\delta({{\bf r}_{ij}})
  +h.c.\}\nonumber\\
& &+t_2(1+x_2P_\sigma)\frac{1}{\hbar^2}{\bf p}_{ij} \cdot \delta({\bf r}_{ij})
 {\bf p}_{ij}\nonumber\\
& &+\frac{1}{6}t_3(1+x_3P_\sigma)\rho^\gamma\delta({\bf r}_{ij})\nonumber\\
& &+\frac{i}{\hbar^2}W_0(\mbox{\boldmath$\sigma_i+\sigma_j$})
\cdot {\bf p}_{ij}\times\delta({\bf r}_{ij}){\bf p}_{ij}
,
\end{eqnarray} 
and a $\delta$-function pairing force acting between like nucleons treated 
either in the full Bogoliubov framework (HFB) \cite{ms02,sg02,ms03}, or the BCS
approximation thereto (HFBCS) \cite{sg01},
\begin{equation}
\label{2}
v_{pair}(\mbox{\boldmath$r$}_{ij})=
V_{\pi q}~\left[1-\eta
\left(\frac{\rho}{\rho_0}\right)^\alpha\right]~
\delta(\mbox{\boldmath$r$}_{ij})
,
\end{equation} 
where $\rho \equiv \rho({\bf r})$ is the local density, and $\rho_0$ is its
equilibrium value in symmetric infinite nuclear matter (INM). 
Only in the most recent paper \cite{ms03} was the possibility of a
density-dependent pairing force admitted; in the first three 
\cite{sg01,ms02,sg02} we had $\eta$ = 0. However, in all four HF mass formulas 
the strength parameter $V_{\pi q}$ was allowed to be different for neutrons 
and protons, and also to be slightly stronger for an odd number of nucleons 
($V_{{\pi q}}^-$) than for an even number ($V_{{\pi q}}^+$), i.e., 
the pairing force between neutrons, for example, depends on whether $N$ is 
even or odd. The two HFB mass formulas \cite{sg02,ms03} add to the energy 
corresponding to the above force 
(and to the kinetic energy and Coulomb energy including the exchange term
in Slater approximation) a phenomenological Wigner term of the form
\begin{eqnarray}
E_W 
& = & V_W\exp\Bigg\{-\lambda\Bigg(\frac{N-Z}{A}\Bigg)^2\Bigg\}
      \nonumber \\
&   & +V_W^{\prime}|N-Z|\exp\Bigg\{-\Bigg(\frac{A}{A_0}\Bigg)^2\Bigg\} 
; 
\label{3}
\end{eqnarray}
a somewhat simpler Wigner term was used in the HFBCS mass formula \cite{sg01}
and the first HFB mass formula \cite{ms02}.

An important question concerns the cutoff to be applied to the 
$\delta$-function pairing force: both BCS and Bogoliubov calculations
diverge if the space of single-particle (s.p.) states
over which such a pairing force is allowed to act is not truncated.
However, making such a cutoff is not simply a computational device but
is rather a vital
part of the physics, pairing being essentially a finite-range phenomenon.
To represent such an interaction by a
$\delta$-function force is thus legitimate only to the extent that all
high-lying excitations are suppressed, although how exactly the truncation of
the pairing space should be made will depend
on the precise nature of the real, finite-range pairing force. It was precisely
our ignorance on this latter point that allowed us in Ref.\ \cite{sg02} to
exploit the cutoff as a new degree of freedom: we found
an optimal mass fit with the spectrum of s.p.\ states $\varepsilon_i$ confined
to lie in the range
\begin{equation}
E_F - \varepsilon_{\Lambda} \le \varepsilon_i \le E_F + \varepsilon_{\Lambda}
,
\label{3A}
\end{equation}
where $E_F$ is the Fermi energy of the nucleus in question, and
$\varepsilon_{\Lambda}$ is a free parameter. We shall adopt the same 
parametrization in the present paper.

The two most recent of our mass formulas \cite{sg02,ms03} had their forces,
labeled BSk2 and BSk3, respectively, fitted to the 2135 nuclei 
with $Z,N \geq 8$ whose masses have been measured and compiled in the 2001 
Atomic Mass Evaluation (AME) of Audi and Wapstra \cite{aw01}. The essential
difference between these two forces is that the pairing is 
density-independent in the case of BSk2 \cite{sg02}, while having a
density dependence of the form (\ref{2}), with the parameters $\eta$ and
$\alpha$ taking the values given by Garrido {\it at al.}\ \cite{gar99},
in the case of BSk3 \cite{ms03} (no other choice for $\eta$ and $\alpha$ leads
to a significant improvement). The rms errors of
these fits are 0.674 MeV \cite{sg02} and 0.656 MeV \cite{ms03}, respectively;
the slight superiority of the latter is too insignificant to imply
that the mass data require the pairing to be density-dependent, and the most
that one can say is that a density dependence of the form of Ref.\  \cite{gar99}
is not ruled out. (On the other hand, the simple model of $\eta$ = 1, 
corresponding to vanishing pairing in the nuclear interior, is quite
incompatible with the mass data.) Using these
two forces, complete mass tables, referred to as HFB-2 \cite{sg02} and HFB-3
\cite{ms03}, respectively, were constructed, including
all the 9200 nuclei lying between the two drip lines over the range of $Z$ and 
$N \ge 8$ and $Z \le 120$. 

This paper is the second in a series of studies of possible
modifications to the original HFB-2 calculation \cite{sg02}, with respect to 
both the force model and the method of calculation. Our motivation for making
such modifications is mainly astrophysical: see Section I of paper I of
this series \cite{ms03}, in which we dealt with the question of the density 
dependence of the pairing force. The main purpose of the present paper is to 
examine the role of the effective nucleon mass in Skyrme-HFB mass formulas. 

We begin by recalling that the HF equation, i.e., the equation determining the 
s.p.\ states, has for the Skyrme forces (\ref{1}) 
and our choice of symmetries the particularly simple form
\begin{eqnarray}
\label{4}
\lefteqn{
\Bigg\{-\mbox{\boldmath$\nabla$}\cdot\frac{\hbar^2}{2M^*_q({\bf r})}
\mbox{\boldmath$\nabla$} + U_q({\bf r}) +V^{coul}_q({\bf r}) 
} \nonumber \\
&  & - i{\bf W}_q({\bf r})\cdot\mbox{\boldmath$\nabla$}
\times\mbox{\boldmath$ \sigma$}\Bigg\}\phi_{i,q}=
\epsilon_{i,q}\phi_{i,q}
.
\end{eqnarray}
All quantities appearing here are defined as in, for example, 
Ref.\ \cite{ton00}, but the essential point is that all the non-locality 
is confined to the effective mass $M^*_q({\bf r})$, which is 
given in terms of the Skyrme parameters by
\beqy\label{5} \frac{\hbar^2}{2M^*_q({\bf r})} = \frac{\hbar^2}{2M} + 
\frac{1}{8}\big\{t_1(2 + x_1) + t_2(2 + x_2)\big\}\rho({\bf r}) \nonumber  \\
+\frac{1}{8}\big\{t_2(2x_2 + 1)-t_1(2x_1 + 1)\big\}\rho_q({\bf r}) \quad ,
\eeqy
the subscript $q$ denoting neutron or proton. At any point in the nucleus the
two effective masses $M^*_n({\bf r})$ and $M^*_p({\bf r})$ are now seen to be
determined entirely by the local densities according to
\beqy\label{6}
\frac{\hbar^2}{2M^*_q} =
\frac{2\rho_q}{\rho}\frac{\hbar^2}{2M^*_s} +
\left(1-\frac{2\rho_q}{\rho}\right)\frac{\hbar^2}{2M^*_v}
,
\eeqy
where $M^*_s$ and $M^*_v$ are the so-called isoscalar and 
isovector effective masses, respectively, quantities that are determined 
by the Skyrme-force parameters according to
\begin{subequations}
\beqy\label{7a}
\frac{\hbar^2}{2M^*_s} = \frac{\hbar^2}{2M} +
\frac{1}{16}\Big\{3t_1+t_2(5+4x_2)\Big\}\rho
,
\eeqy
and
\beqy\label{7b}
\frac{\hbar^2}{2M^*_v} = \frac{\hbar^2}{2M} + 
\frac{1}{8}\Big\{t_1(2+x_1)+t_2(2+x_2)\Big\}\rho
.
\eeqy
\end{subequations}

The single-particle energies $\epsilon_{i,q}$ of even nuclei, obtained as 
eigenvalues of the HF Hamiltonian, Eq.\ (\ref{4}), are often identified with 
the  one-nucleon separation energies into or from certain low-lying excited 
states in adjacent odd-$A$ nuclei, see \cite{be03} and references 
therein. 
It is known that to have the same density of s.p.\ levels 
$\epsilon_{i,q}$ in the vicinity of the Fermi level as observed in
experiment for heavy and intermediate-mass nuclei, one must have 
$M^*_s/M$ equal to, or close to, 1.0 at saturation density $\rho_0$ 
\cite{bro63,bt81} (we see from Eq.\ (\ref{6}) that the isovector effective 
mass $M_v$ will have little influence on s.p.\ energies of nuclei that are
relatively close to the stability line). On the other hand, INM calculations 
with forces that are realistic in the sense that they fit the two- and 
three-nucleon data (and therefore require an explicit treatment of 
the short-range correlations that are built in an effective way into 
the forces for HFB calculations)
indicate that $M^*_s/M$ lies in the range
0.6--0.9 for $\rho = \rho_0$ \cite{bg58,fp81,wir88,zuo99,zuo02}. Rough 
experimental confirmation that $M^*_s$ is indeed significantly smaller than $M$
came first from measurements of the deepest s.p.\ states in light nuclei 
\cite{jam69} (the deepest s.p.\ states of heavier nuclei have not been 
measured); see Refs.\ \cite{vb72,jlm76,mah85} for theoretical discussions. More 
precise empirical information comes from analyses of the giant isoscalar 
quadrupole resonance, which lead to a value of around 0.8$M$ for $M^*_s$ at 
$\rho = \rho_0$, according to Ref.\ \cite{boh79}. 

Actually, there is no contradiction between these two sets of values of
$M^*_s/M$, since Refs.\ \cite{bk68,bg80} have shown that in finite nuclei 
one can obtain reasonable s.p.\ level densities near the Fermi level 
with the INM values of $M^*_s/M$, i.e., of 0.6--0.9, provided one takes into
account the coupling between s.p.\ excitation modes and surface-vibration RPA
modes. Since the good agreement with measured s.p.\ level densities found in
Ref.\ \cite{bt81} was obtained without making these corrections it must be
supposed that the resulting error is being compensated by the higher value of
$M^*_s/M$, i.e., $M^*_s/M \simeq 1.0$, which may thus be regarded as a
phenomenological value that leads to good agreement with measured s.p.
energies in straightforward
HF, or other mean-field calculations, without any of the complications of
Refs.\ \cite{bk68,bg80}.

Now the fact that in all our previous mass fits \cite{sg01,ms02,sg02,ms03}, 
where no constraints were placed on the effective mass, we found 
$M^*_s/M \simeq 1.0$ at the density $\rho = \rho_0$ 
suggests that obtaining a correct 
s.p.\ spectrum in the vicinity of the Fermi level is a necessary condition for
an optimum mass fit. This conclusion tends to be confirmed by the 
occupation-number representation of the Strutinsky theorem, which approximates
the total energy of the nucleus as
\beqy\label{8}
E \simeq \tilde{E} + \sum_{i}\epsilon_i\,\delta n_i \quad ,
\eeqy
where $\tilde{E}$ is a smoothed, average value of $E$, while $\delta\,n_i =
n_i - \tilde{n_i}$, in which $n_i$ is the actual occupation number of the
s.p.\ state $i$ and $\tilde{n_i}$ is an average occupation number, given by, 
for example, Eq.\ (IV.18) of the review of Brack {\it et al.}\ \cite{brack72}. 
But in Section IV.6 of this same paper \cite{brack72} it is shown that 
$\delta n_i$ is non-vanishing only for s.p.\ states lying within about
20$A^{-1/3}$ MeV of the Fermi level. We may expect that it will be difficult to
obtain correct masses if the s.p.\ spectrum over this interval is not
reproduced, and it is quite comprehensible that the optimal mass fits published
so far require that $M^*_s/M$ take a value close to 1.0 at the density 
$\rho = \rho_0$. Indeed, with a Skyrme force of the form (\ref{1}) and
density-independent $\delta$-function pairing forces treated within the BCS
approximation, Farine {\it et al.} \cite{far01} found that if
$M_s^*/M$ was constrained to be equal to 0.8 then in fitting 416
quasi-spherical nuclei it was impossible to reduce the rms error below
1.141 MeV (parameter set MSk5$^*$). On the other hand, when the constraint on
$M_s^*/M$ was released the rms error for the same data set fell to 0.709 MeV,
$M_s^*/M$ rising to 1.05 (parameter set MSk5). Moreover, the Skyrme force 
SLy4 \cite{cha98} with $M_s^*/M$ = 0.7 
does not reproduce very well the masses of the few open-shell
nuclei considered (see Figs. 1--4 of Ref. \cite{cha98}). 

Nevertheless, there remains some room for maneuver in Eq. (\ref{8}), since 
shifts in the s.p.\ energies $\epsilon_i$ resulting from a change
in the effective mass could in principle be compensated by appropriate changes
in the $\delta\,n_i$, provided full use was made of all the degrees of freedom 
in the force, with particular emphasis
on those that have not hitherto been exploited; of  especial interest in this
respect is the pairing cutoff, which, for example, was fixed at the same value 
in the MSk5 and MSk5$^*$ fits (41~$A^{-1/3}$ MeV into the continuum).
In the present paper we pursue these possibilities with the aim of seeing to 
what extent it is possible to maintain quality mass fits with lower values of 
$M^*_s/M$, more appropriate to INM. 
Part of our own interest in this question lies in our search for
a unique effective interaction suitable for the determination of an equation of
state describing the formation of nuclear matter from isolated finite 
nuclei that occurs during stellar collapse \cite{far01,op02}. 
A lower value of $M^*_s/M$, corresponding to INM, 
will certainly be appropriate in the final 
stage; the question here is to see to what extent such a choice of $M^*_s/M$ 
can suitably describe the isolated nuclei prevailing at the beginning of the 
collapse. (Similar considerations will arise in the more or less inverse 
sequence of events traced out during the neutron-matter decompression that 
occurs, for example, in the aftermath of neutron-star mergers.)  

In this paper
we also replace our original approximate correction for the centre-of-mass
motion by a much improved treatment, and we describe this first, in
Section II, leaving until Section III the discussion of the role of the
effective mass. There, two different values of $M^*_s$ at the density 
$\rho = \rho_0$, 0.92$M$ and 0.8$M$, will be considered, and four new mass
tables, HFB-4, HFB-5, HFB-6, and HFB-7, generated,  each value of $M^*_s$ being
calculated with and without a density dependence in the pairing force. 

%******************************************
\section{Center-of-mass correction}
\label{sect_cm}
%******************************************

Mean-field approaches like HF or HFB, which establish an intrinsic frame 
of the nucleus, break several symmetries of the Hamiltonian and the wave 
function in the laboratory frame \cite{rs80,be03}. For example, finite 
nuclei break translational invariance, deformed nuclei are not rotational 
invariant, and the HFB approach breaks particle-number symmetry.
Doing so adds desired correlations to the modeling -- as
multi-particle-multi-hole states associated with deformation, pairing,
etc. -- but at the same time gives rise to an admixture of excited 
states to the calculated ground state. Their spurious contribution 
to the total mass changes with nucleon numbers and deformation. The 
energy of those spurious modes, which are not explicitly removed from
the calculated ground state, will be simulated by the Skyrme force
through the parameter fit, which might spoil the properties of the
resulting forces \cite{be00}.

A rigorous way to restore the broken symmetries is projection on 
exact quantum numbers, but this would be too time-consuming
to be used for the large-scale mass fits performed here. A simpler 
procedure is to estimate the contribution to the binding energy in a
suitable approximation, and to add the resulting corrections to the
calculated masses; such a procedure has already been adopted by many workers
for the center-of-mass (cm) correction, the rotational correction, and the 
Lipkin-Nogami correction to the pairing energy. In our own HF mass formulas 
\cite{sg01,ms02,sg02,ms03} we have used such a procedure for the
c.m. and rotational corrections, as described in Ref. \cite{ton00}.
In the present paper we improve our treatment of the cm correction, as 
discussed below in this section, but otherwise the Skyrme-HFB formalism
used here is essentially as described in detail in our
first HFB paper \cite{ms02}. In particular, we have not yet made any correction
for particle-number fluctuations; this will be the topic of a forthcoming 
paper.

The HFB ground state is not an eigenstate of the total momentum operator.
Thus, although the expectation value of the momentum operator
$\hat{\bf P} \equiv \sum_i \hat{\bf p}_i$ in the cm frame
$\langle {\rm HFB} |\hat{\bf P}| {\rm HFB}\rangle$ vanishes, its
dispersion $\langle {\rm HFB}|\hat{\bf P}^2| {\rm HFB}\rangle$ does not.
Gaussian overlap approximation to exact momentum projection
gives for the spurious cm energy
\beqy\label{cm.1}
E_{cm} =
\frac{1}{2MA}\langle {\rm HFB}|\hat{\bf P}^2| {\rm HFB}\rangle
,
\eeqy
which has to be subtracted from the calculated total energy. 
To avoid the time-consuming evaluation of the non-diagonal 
terms ${\bf \hat{p}}_i \cdot {\bf \hat{p}}_j, i\ne j$,
in the past we have always adopted the approximation of 
Butler {\it et al.}\ \cite{butler}, which takes explicit 
account only of the diagonal terms, thus
\beqy\label{cm.2}
E_{cm} \simeq \frac{f(A)}{2MA}\langle {\rm HFB}|
\sum_i \hat{\bf p}_i^2| {\rm HFB}\rangle
\quad ,
\eeqy
where $f(A)$ is a simple function that makes this expression exact in the case
of pure oscillator s.p.\ states.
From now on, we evaluate the cm correction according to Eq.\ (\ref{cm.1}), 
doing so, however, perturbatively. That is, both the diagonal and 
off-diagonal terms of Eq.\ (\ref{cm.1}) are included only in the calculation of 
the converged total energy, not in the variational equation that 
leads to the mean field in the HF equation (\ref{4}).

\begin{figure}
\centerline{\epsfig{figure=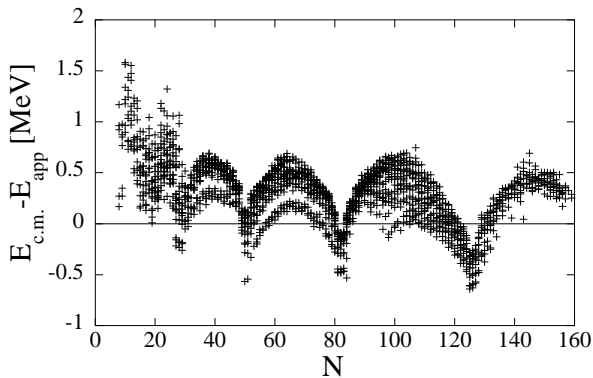,width=8cm}}
\caption{Comparison for the 2135 nuclei included in the 2001 mass compilation 
of \cite{aw01} of the binding energies obtained with the improved 
centre-of-mass correction $E_{cm}$ with those obtained with the approximation 
of \cite{butler} $E_{app}$. Both calculations use the BSk2 Skyrme force and are 
made in the spherical approximation.}
\label{fig_cm}
\end{figure}

The effect of this improved treatment is seen in Fig.~\ref{fig_cm}, where
for each of the 2135 nuclei of known mass we show the difference between the
total energy calculated with our new method, and that calculated with the 
approximation of 
Ref.\ \cite{butler}. The force used for this comparison is BSk2, and for 
simplicity we assume a spherical configuration for all nuclei. The differences 
are largest for light nuclei, for which they can reach 1.5 MeV; strong shell 
effects will be noticed.  

\begin{figure}
\centerline{\epsfig{figure=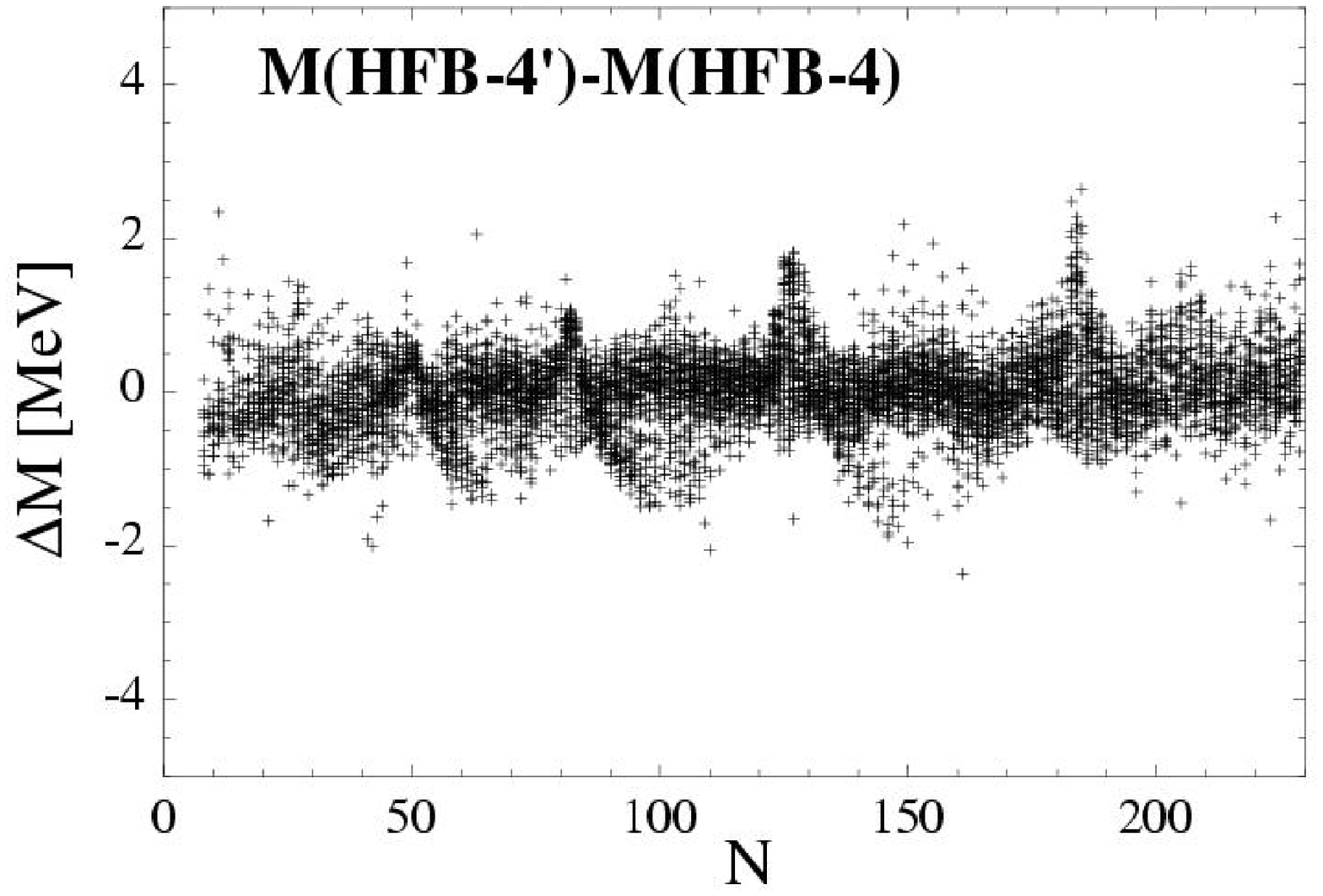,width=8cm}}
\caption{Differences between the HFB-4$^{\prime}$ and HFB-4 masses
as a function of the neutron number $N$ for all nuclei with $Z, N \ge$ 8
lying between the proton and neutron driplines up to $Z = 120$.}
\label{fig_hfb4p}
\end{figure}

Of course, in fitting force BSk2 to the mass data these errors in the 
approximation of Ref.~\cite{butler} were absorbed to some extent into the 
force parameters. It would be interesting to refit the BSk2 force to the same 
data using the improved cm correction, and compare the new with the original
BSk2 fit. For practical reasons, we make this comparison on the basis of the
force BSk4, described in the next section. This force (like the forces BSk5,
BSk6 and BSk7) is calculated with the improved cm correction \cite{be00}, so
we refit the force BSk4 to the data using the approximate cm correction of
Ref.~\cite{butler}, defining thereby the force BSk4$^{\prime}$. We find that
the change in the rms error of the fit is negligible, going from 0.680
to 0.681 MeV. The mass differences between the corresponding BSk4$^{\prime}$ 
and BSk4 predictions are compared in Fig.~\ref{fig_hfb4p} for all nuclei with 
$Z, N \ge$ 8
lying between the proton and neutron driplines up to $Z = 120$. In contrast to
Fig.~\ref{fig_cm}, deformation effects are taken into account consistently. The
difference in the shell structure between the improved treatment of the 
cm correction  \cite{be00} and the approximation of Ref.~\cite{butler}, as 
seen in Fig.~\ref{fig_cm}, is still present after renormalizing the Skyrme 
forces on experimental masses. The mass differences remain however smaller than
2~MeV, even for exotic neutron-rich or superheavy nuclei.

 \begin{table*}[t!]
 \caption{Rms ($\sigma$) and mean ($\bar{\epsilon}$) errors (in MeV) in the
predictions of masses $M$ obtained with the BSk2-7 forces. Also given are the 
model standard deviation and the model mean error (see text for more details) 
on the set of all 2135 measured masses, on the 928 stable and neutron-rich 
nuclei, and on the 1207 proton-rich nuclei. The last two lines correspond to 
the rms and mean errors (in fm) for the predictions of the 523 measured charge 
radii ($r_c$).}
\label{tab_rms}
 \tabcolsep=.3cm
 \begin{tabular}{|c|c|c|c|c|c|c|}
 \hline
  & BSk2 & BSk3 & BSk4  & BSk5 & BSk6 & BSk7 \\
 \hline
 $\sigma(M)$ (2135 nuclei) & 0.674 & 0.656 & 0.680   & 0.675  & 0.686 & 0.676 \\
  $\bar{\epsilon}(M)$ (2135 nuclei) & 0.000 & -0.006 & -0.115  & -0.005  & -0.013 & -0.004 \\
 $\sigma_{mod}(M)$ (2135 nuclei)& 0.660 & 0.639 & 0.661   & 0.655  & 0.666 & 0.658 \\
 $\bar{\epsilon}_{mod}(M)$ (2135 nuclei) & -0.007 & -0.015 & 0.106   & -0.006 & 0.013 & 0.026\\
 $\sigma_{mod}(M)$ (928 nuclei)& 0.709 & 0.709 & 0.678   & 0.685  & 0.713 & 0.707 \\
 $\bar{\epsilon}_{mod}(M)$(928 nuclei)& -0.168 & -0.118 & 0.001   & -0.122 & 0.093 & -0.085\\
 $\sigma_{mod}(M)$ (1207 nuclei)& 0.620 & 0.581 & 0.649   & 0.631  & 0.629 & 0.618 \\
 $\bar{\epsilon}_{mod}(M)$(1207 nuclei)& 0.113 & 0.062 & 0.186   & 0.080 & 0.092 & 0.109\\
 \hline
 $\sigma(r_c)$ (523 nuclei)       & 0.0282 & 0.0291 & 0.0282  & 0.0270 & 0.0262  &
0.0260 \\
  $\bar{\epsilon}(r_c)$ (523 nuclei) & 0.0138 & 0.0161 & -0.0115  & 0.0104 &0.0028 &
-0.0030 \\
 \hline
 \end{tabular}
 \end{table*}

\begin{table*}[t!]
\caption{Skyrme-force and pairing-force parameters of BSk2-BSk7}
\label{tab_sky}
 \tabcolsep=.2cm
 \begin{tabular}{|ccccccc|}
 \hline
  & BSk2 & BSk3 & BSk4 & BSk5 & BSk6 & BSk7 \\
 \hline
  $t_0$ {\scriptsize [MeV fm$^3$]} & -1790.6248 & -1755.1297      & -1776.9376 & -1778.8934
 & -2043.3174 & -2044.2484 \\
  $t_1$ {\scriptsize [MeV fm$^5$]} & 260.996& 233.262      & 306.884 & 312.727 & 382.127 
& 385.973 \\
  $t_2$ {\scriptsize [MeV fm$^5$]} & -147.167& -135.284 & -105.670 & -102.883 &
-173.879 & -131.525 \\
  $t_3$ {\scriptsize [MeV fm$^{3+3\gamma}$]} & 13215.1&13543.2 & 12302.1 & 12318.37 &
12511.7 & 12518.8 \\
  $x_0$             & 0.498986 &  0.476585      &  0.542594 & 0.444510 & 0.735859 &
0.729193\\
  $x_1$             & -0.089752   & -0.032567      & -0.535165 & -0.488716 & -0.799153 & 
-0.932335\\
  $x_2$  & 0.224411  &  0.470393     &  0.494738 & 0.584590 &-0.358983 & -0.050127\\
  $x_3$             & 0.515675  &  0.422501    &   0.759028 & 0.569304 & 1.234779 &
.236280\\
  $W_0$ {\scriptsize [MeV fm$^5$]} & 119.047& 116.07  & 129.50 & 130.70 & 142.38 & 146.93\\
  $\gamma$          & 0.343295   & 0.361194 & 1/3 & 1/3 & 1/4 & 1/4\\
  $V^+_n$ {\scriptsize [MeV fm$^3$]} & -238 & -359  & -273 & -429 & -321 & -505\\

  $V^-_n$ {\scriptsize [MeV fm$^3$]} & -265 &-407  & -289 & -463 & -325 & -514 \\

  $V^+_p$ {\scriptsize [MeV fm$^3$]} & -247 &-365  & -285 & -447 & -338 & -531\\

  $V^-_p$ {\scriptsize [MeV fm$^3$]} & -278 &-413  & -302 & -483 & -341 & -541\\

  $\eta$                      & 0 & 0.45 & 0 & 0.45 & 0 & 0.45\\
  $\alpha$                    & 0  & 0.47 & 0  & 0.47 & 0 & 0.47 \\
  $\varepsilon_{\Lambda}${\scriptsize [MeV]} & 15 & 14 & 16  & 16 & 17 & 17\\
  $V_W$ {\scriptsize [MeV]} & -2.05 &-2.05  & -1.72 & -1.96 & 1.76 & 1.86\\
  $\lambda$ & 485 & 460  & 740 & 480 & 700 & 720\\
  $V_W^{\prime}$ {\scriptsize [MeV]} & 0.70 & 0.54  & 0.54 & 0.50 & 0.58 & 0.54\\
  $A_0$ & 28 & 30  & 30 & 30 & 28 & 28\\
 \hline
 \end{tabular}
 \end{table*}

\section{Choice of the isoscalar effective mass, and the new HFB mass tables}
\label{sect_mass}

Most INM calculations of $M_s^*$ at the density $\rho = \rho_0$ give a value of
around 0.8$M$, the most recent such calculation being that of Ref.\ 
\cite{zuo02}.  On the other hand, using the so-called extended
Brueckner-Hartree-Fock method with realistic nucleonic forces,  
Ref.\ \cite{zuo99} finds 0.92$M$.
Rather than attempt to decide between these two values we shall here consider
both of them, with the value of 0.92$M$ being imposed on parameter sets
BSk4 and BSk5 (mass formulas HFB-4 and HFB-5, respectively), and the value of
0.8$M$  being imposed on parameter sets BSk6 and BSk7 (mass formulas HFB-6 and
HFB-7, respectively). The pairing force in both BSk4 and BSk6 is supposed to
be density-independent, while that of BSk5 and BSk7 is taken to have a
density dependence of the form (\ref{2}), with the parameters $\eta$ and
$\alpha$ taking the values given by Garrido {\it at al}.\ \cite{gar99}, as with
the BSk3 force \cite{ms03}. In all four cases the isovector mass at 
$\rho = \rho_0$ is left unconstrained in the fits.

We fit all four of these forces to the same data set as the one to which
BSk2 \cite{sg02} and BSk3 \cite{ms03} were fitted, i.e., the 2135 nuclei
with $Z,N \geq 8$ whose masses have been measured and compiled in the 2001
AME \cite{aw01}. 
As in the fits of BSk2 and BSk3, we impose a lower
limit on the INM symmetry coefficient $J$, in order to prevent the collapse of
neutron matter at nuclear densities, as required by the observed stability
of neutron stars.  In the case of BSk5, we use the INM calculation 
of \cite{zuo99} to constrain not only the effective mass to 0.92$M$, 
but also the symmetry coefficient to $J=28.7$~MeV. Also, we set the 
equilibrium density of symmetric INM at $\rho_0=0.1575$~fm$^{-3}$, it 
having been found that this ensures very good predictions for
rms charge radii along with near-optimal mass fits.

\begin{figure}[t!]
\centerline{\epsfig{figure=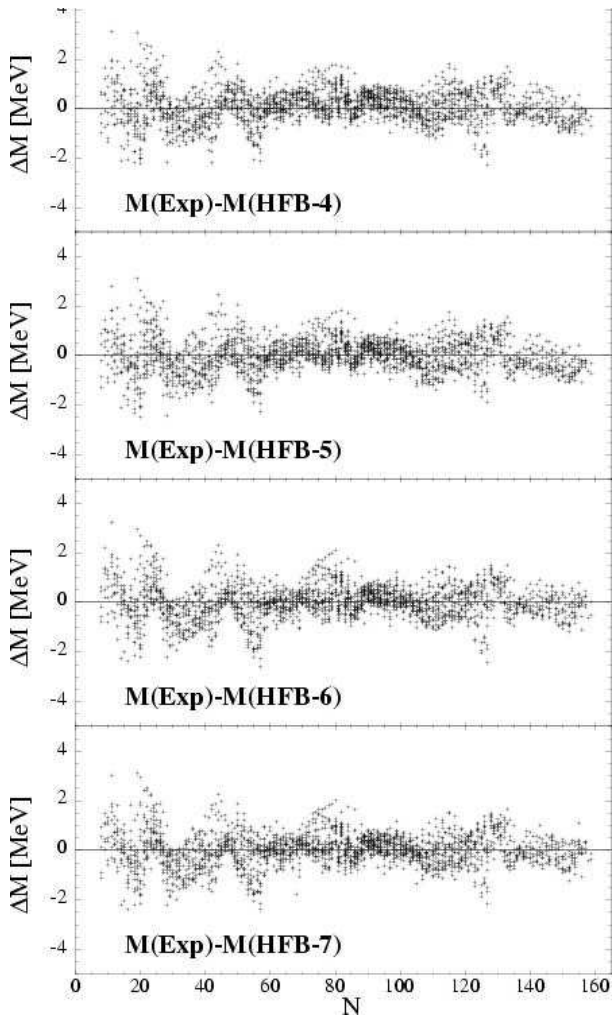,width=8cm}}
\caption{Differences between experimental and calculated mass excesses as a 
function of the neutron number $N$ for the HFB-4 to HFB-7 mass tables. }
\label{fig_exp}
\end{figure}

From the first line of Table \ref{tab_rms} we see that while forces BSk2 and
BSk3, in which $M_s^*$ is unconstrained, still give the best fits, all four of 
the new forces, BSk4--7, give fits that are almost as good, the deterioration
on constraining $M_s^*$ being quite negligible. Fig. \ref{fig_exp} displays the
deviations of the calculated masses from the experimental values.

Table~\ref{tab_rms} also has entries for a ``model standard deviation"
$\sigma_{mod}$ and a ``model mean error" $\bar{\epsilon}_{mod}$.
These quantities have been introduced \cite{mn88,frdm} as an improved measure
of the validity of the physical model that is being fitted to the data, the
standard rms error $\sigma$ suffering from the defect that it does not
take account of the experimental errors of the individual mass measurements, as
given by Audi and Wapstra \cite{aw95,aw01}.
The standard rms error is a legitimate measure when the experimental errors
are small compared to the rms error itself, but some of the most recent
measurements of nuclei far from the stability line  
have errors in excess of 1 MeV \cite{aw01}. In such cases, 
$\sigma_{mod}$ and $\bar{\epsilon}_{mod}$ give a better assessment of the 
validity of a given mass formula, since they weight each data point in terms
of its experimental error, following a procedure based on the method of
maximum likelihood \cite{mn88,frdm}. The definition of $\sigma_{mod}$ that
we adopt is that of Eqns. (42) and (43) of Ref.\ \cite{mn88} (which writes
$\sigma_{mod}$ as $\sigma_{th}^*$), while our definition of
$\bar{\epsilon}_{mod}$ is that of Eqns. (9) and (10) of Ref.\ \cite{frdm} (which
writes $\bar{\epsilon}_{mod}$ as $\mu_{th}^*$); for a discussion of the
relationship between the ways in which the two papers \cite{mn88} and  
\cite{frdm} treat model errors see Appendix B of Ref. \cite{rmp03}. 
We show $\sigma_{mod}$ and $\bar{\epsilon}_{mod}$ for the full data set
of 2135 masses to which the fit was made, as well as to the two subsets
of 1207 proton-rich nuclei and 928 stable and neutron-rich nuclei.  
None of these model errors suggests that any particular force is significantly
better or worse than any of the others, as far as masses are concerned. 
The fact that mass fits with values of $M_s^*/M$ constrained to be equal to 0.8
can be obtained which are almost as good as those for which $M_s^*/M$ is
unconstrained (always emerging with a value close to 1.0) is, we have found,
essentially a consequence of our exploitation of the degree of freedom
associated with the pairing cutoff, a reduction in $M_s^*/M$ being almost
completely compensated by an increase in the parameter $\varepsilon_{\Lambda}$
appearing in Eq.\ (\ref{3A}) (see Table \ref{tab_sky}).

 \begin{table}[b!]
 \centering
 \caption{Macroscopic parameters of the forces BSk2-BSk7}
\label{tab_par}
\tabcolsep=.1cm
 \begin{tabular}{|c|cccccc|}
 \hline
  & BSk2 & BSk3 & BSk4 & BSk5 & BSk6 & BSk7 \\
 \hline
  $a_v$ {\scriptsize [MeV]} & -15.794 & -15.804    & -15.773 & -15.802 & -15.749 & -15.760\\
  $\rho_0$ {\scriptsize [fm$^{-3}$]} & 0.1575 & 0.1575 & 0.1575 & 0.1575 & 0.1575 & 0.1575\\
  $J$ {\scriptsize [MeV]} & 28.0 & 27.9 & 28.0 & 28.7 & 28.0 & 28.0 \\
  $M^*_s/M$ & 1.04 & 1.12 & 0.92 & 0.92 & 0.80 & 0.80 \\
  $M^*_v/M$ &0.86 & 0.89 & 0.85 & 0.84 & 0.86 & 0.87 \\
  $K_v$ \scriptsize [MeV] &233.6  & 234.8   & 236.8 & 237.2 & 229.1 & 229.3\\
  $G_0$     & -0.705  &  -0.994     &  -0.478  & -0.579 & 0.065 & -0.101 \\
  $G_0^{'}$  &0.446   &   0.497      &  0.457   & 0.454 & 0.312 & 0.356 \\
  $\rho_{frmg}/\rho_0$   & 1.1  & 1.00 & 1.30  & 1.20 & 1.81 & 1.62 \\
  $a_{sf}$ {\scriptsize [MeV]} & 17.5 & 17.5&17.3 &17.5&17.2&17.3\\
  $Q$ {\scriptsize [MeV]} & 68 & 67&76&52&83&80\\
 \hline
 \end{tabular}
 \end{table}

We also show in Table~\ref{tab_rms} the rms and mean deviations 
between our calculated 
and experimental charge radii for the 523 nuclei listed in the 1994 compilation
\cite{nad94} (for more details on the HFB derivation of the charge radii, see
Ref. \cite{bpg01}). The overall agreement with experiment is seen to be 
excellent. However, none of the forces is able to completely 
reproduce the much discussed kink in the Pb isotope chain at $N$ = 126.

\begin{figure}[b!]
\centerline{\epsfig{figure=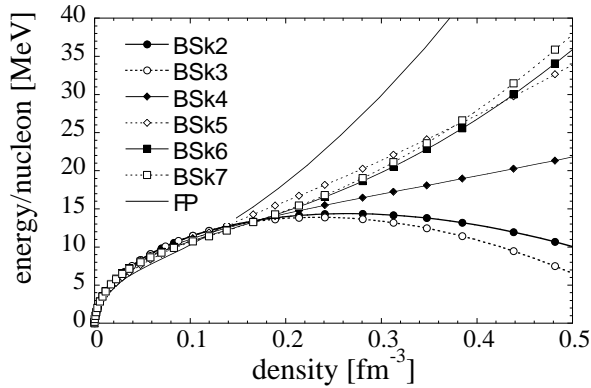,width=8cm}}
\caption{Energy-density curves of neutron matter for the forces
of this paper, and for the calculations of Ref.\ \cite{fp81} (FP). }
\label{fig_nm}
\end{figure}

The parameters of the forces BSk2--7 are given in Table~\ref{tab_sky},
while the corresponding macroscopic parameters, i.e., the parameters relating 
to INM and SINM (semi-infinite nuclear matter) calculated for all these forces,
are shown in Table~\ref{tab_par}. The quantities appearing in this latter
table that have not yet been defined are: $a_v$, the energy per nucleon at 
equilibrium in symmetric INM; $K_v$, the INM incompressibility;
$G_0$ and $G_0^{\prime}$, the Landau parameters defined in Ref.\ \cite{gs81};  
$\rho_{frmg}$, the density at which neutron matter flips over into a 
ferromagnetic state that has no energy minimum and would collapse indefinitely 
\cite{kw94}; $a_{sf}$, the surface coefficient; and $Q$, the surface-stiffness 
coefficient \cite{ms69}. It will be recalled that in all cases the values of
$\rho_0$ and $J$ were imposed, as were the values of $M_s^*/M$. On the
other hand, $M_v^*/M$ is unconstrained, but it is gratifying to note that
for all fits its value comes out to be quite close to 
the value of 0.83 that we infer  from the INM calculations
of Zuo {\it et al.}\ \cite{zuo99} (see especially their Fig. 9).
Nevertheless, a word of caution is necessary here, since it is known from 
Ref.\ \cite{pg01} that the rms error of the mass fit varies only slowly with
$M^*_v$. As for $K_v$, all our forces give values falling within the 
experimental
range of 225--240 MeV established by Youngblood {\it et al.}\ \cite{ylc02}. 
All our forces likewise satisfy the condition $G_0$ and $G_0^{\prime} > -1$
for stability against spin and spin-isospin flips \cite{bjs75} 
at saturation density. 
Finally, it will be seen that reducing $M_s^*$ seems to 
assure greater stability to neutron matter against ferromagnetic flips. 
For the values of $\rho_{frmg}$, $G_0$ and $G_0^{\prime}$ given in 
Table~\ref{tab_par}, it is assumed that the effective spin-spin 
interaction is obtained from the exchange terms of the two-body 
Skyrme force, eq.\ (\ref{1}). There is another possible view of 
the effective Skyrme interaction, see \cite{be03,be02} and references
given therein, which leaves the coupling constants that determine 
$\rho_{frmg}$, $G_0$ and $G_0^{\prime}$ as additional free parameters 
not constrained by our current mass fit.

{\it Neutron Matter.} All our present forces have a value of $J \simeq 28$~MeV
(except BSk5 with $J=28.7$~MeV) which was actually set as a lower limit in the
search on the Skyrme parameters. It is possible that a slightly better mass fit
could have been 
obtained with a value somewhat closer to 27.5 MeV, but this might have 
engendered an unphysical collapse of neutron matter at nuclear densities. The
situation in neutron matter for our forces is as shown in Fig.~\ref{fig_nm};
the solid curve labeled FP  shows the results of Friedman
and Pandharipande \cite{fp81} for the realistic force
Argonne v$_{14}$ + TNI,
containing two- and three-nucleon terms. More recent realistic calculations
of neutron matter \cite{cdl87,wff88,apr98} give similar results up to
nuclear densities. At low densities, all forces give the same predictions. 
At densities higher than the nuclear density, where the validity of the 
non-relativistic approach followed here could admittedly be questioned, the new
forces are found to lead to harder neutron-matter curves and avoid the collapse
obtained for BSk2.  In particular, BSk5 with $J=28.7$~MeV gives very similar
energies per nucleon as BSk6 and BSk7 with  $J=28$~MeV due to the much
lower value of the $x_1$
parameter reached by the mass fit (see Table~\ref{tab_sky}).  
 
\begin{table}[t!]
 \centering
 \caption{Single-particle proton levels in $^{208}$Pb (MeV).  
Experimental values are taken from Ref.\ \cite{vb72}. The asterisk denotes
the Fermi level. The quantity $\Delta_p$ is the interval between the centroids
of the $1g$ and $2f$ doublets.}
\label{tab_sp1}
\begin{center}
\begin{tabular}{|c|cccccc|c|}
\hline
Level       &  BSk2 &  BSk3 &  BSk6 &  MSk5 &  MSk5$^*$ &SLy4&  Expt.\\
\hline
$1s_{1/2}$  & -33.1 & -31.4 & -38.9 & -31.8 & -40.5 &-44.0& -\\
\ldots & \ldots & \ldots & \ldots & \ldots & \ldots & \ldots & \ldots\\
$1g_{9/2}$  & -14.8 & -14.4 & -16.2 & -14.6 & -16.7 & -17.7&-15.4\\
$1g_{7/2}$  & -11.4 & -11.1 & -12.6 & -11.3 & -13.1 & -13.5&-11.4\\
$2d_{5/2}$  &  -9.8 &  -9.6 & -10.4 & -9.7 & -10.5 & -11.5&-9.7\\
$1h_{11/2}$ &  -8.8 &  -8.7 &  -9.0 &  -8.7 &  -9.2 & -9.7&-9.4\\
$2d_{3/2}$  &  -8.2 &  -8.1 &  -8.5 &  -8.2 &  -8.8 & -9.6&-8.4\\
$3s_{1/2}$* &  -7.6 &  -7.5 &  -7.9 &  -7.6 &  -7.9 & -8.8&-8.0\\
$1h_{9/2}$  &  -3.9 &  -3.9 &  -3.8 &  -4.0 &  -4.1 & -3.8&-3.8\\
$2f_{7/2}$  &  -3.1 &  -3.3 &  -2.6 &  -3.3 &  -2.3 & -2.9&-2.9\\
$1i_{13/2}$ &  -2.2 &  -2.4 &  -1.5 &  -2.4 &  -1.3 & -1.5&-2.2\\
$3p_{3/2}$  &  -0.3 &  -0.6 &   0.6 &   -0.6 &   0.9 & 0.4&-1.0\\
$2f_{5/2}$  &  -0.9 &  -1.1 &   0.0 &  -1.1 &  0.0 & -0.4&-0.5\\
\hline
$\Delta_p$ &11.0  &10.5&13.1&10.7  & 13.9 & 14.0&11.7\\
\hline
\end{tabular}
\end{center}
\end{table}

\begin{table}[t!]
 \centering
 \caption{Single-particle neutron levels in $^{208}$Pb (MeV). 
Experimental values are taken from Ref.\ \cite{vb72}. The asterisk denotes
the Fermi level. The quantity $\Delta_n$ is the interval between the centroids 
of the $2f$ and $3d$ doublets.} 
\label{tab_sp2}
\begin{center}
\begin{tabular}{|c|cccccc|c|}
\hline
Level       &  BSk2 &  BSk3 &  BSk6 &  MSk5 & MSk5$^*$ &SLy4& Expt.\\
\hline
$1s_{1/2}$  & -39.5 & -36.9 & -51.5 & -40.7 & -49.9 & -57.9&- \\
\ldots & \ldots & \ldots & \ldots & \ldots & \ldots & \ldots& \ldots\\
$1h_{9/2}$  & -10.6 & -10.2 & -12.4 & -10.8 & -12.3 & -12.5&-10.9\\
$2f_{7/2}$  & -10.8 & -10.6 & -11.7 & -10.8 & -11.5 & -12.0&-9.7\\
$1i_{13/2}$ &  -9.2 &  -9.1 &  -9.6 &  -9.1 &  -9.4 & -9.6&-9.0\\
$3p_{3/2}$  &  -8.4 &  -8.3 &  -8.9 &  -8.5 &  -8.7 & -9.2&-8.3\\
$2f_{5/2}$  &  -8.2 &  -8.1 &  -8.8 &  -8.3 &  -8.7 & -9.1&-8.0\\
$3p_{1/2}$* &  -7.5 &  -7.4 &  -7.8 &  -7.6 &  -7.7 & -8.1&-7.4\\
$2g_{9/2}$  &  -4.2 &  -4.3 &  -3.6 &  -4.1 &  -3.4 & -3.2&-3.9\\
$1i_{11/2}$ &  -2.6 &  -2.7 &  -2.4 &  -2.7 &  -2.6 & -1.7&-3.2\\
-$1j_{15/2}$ &  -2.4 &  -2.6 &  -1.3 &  -2.1 &  -1.2 & -0.6&-2.5\\
$3d_{5/2}$  &  -1.9 &  -2.1 &  -1.2 &  -1.8 &  -1.0 & -0.7&-2.4\\
$4s_{1/2}$  &  -1.2 &  -1.4 &  -0.6 &  -1.0 &  -0.3 & 0.0&-1.9\\
$2g_{7/2}$  &  -1.2 &  -1.4 &  -0.2 &  -1.1 &  -0.3 & 0.0&-1.5\\
$3d_{3/2}$  &  -0.8 &  -1.0 &  0.0 &  -0.7 &  0.1 & 0.3&-1.4\\ 
\hline
$\Delta_n$  &8.2  &7.8  &9.8 &8.3 &9.7 &10.5& 7.2\\
\hline
\end{tabular}
\end{center}
\end{table}

\begin{figure*}[t!]
\centerline{\epsfig{figure=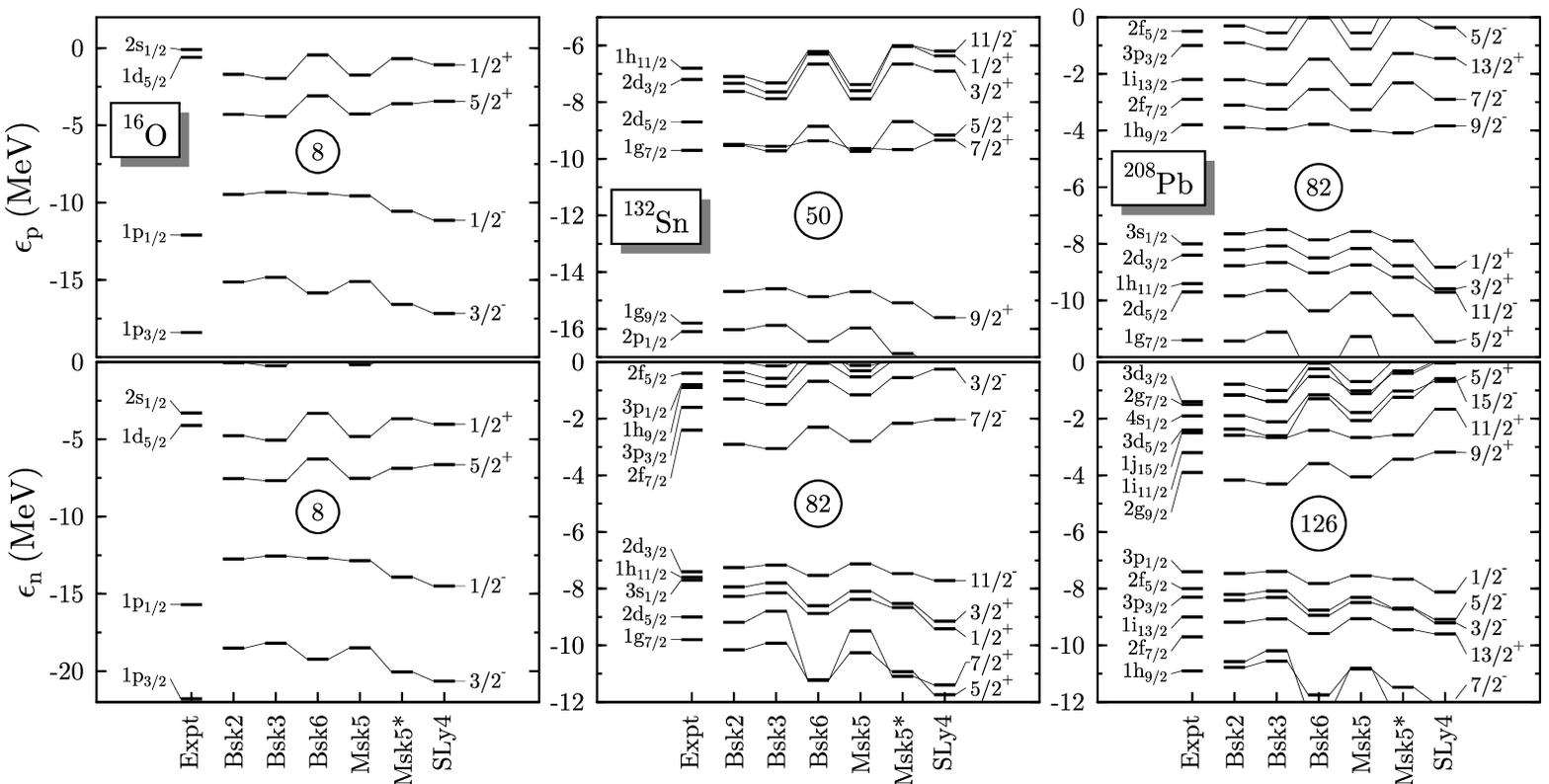,width=16cm}}
\caption{Single-particle spectra for $^{16}$O, $^{132}$Sn, and $^{208}$Pb.
}
\label{fig_sp}
\end{figure*}

\begin{table}[b!]
 \centering
\caption{Single-particle proton levels in $^{132}$Sn (MeV).
Experimental values are taken from Refs.\ \cite{san99,mez95}. The asterisk denotes
the Fermi level. The quantity $\Delta_p$ is the interval between the centroids 
of the $1g$ and $2d$ doublets.} 
\label{tab_sp3}
\begin{center}
\begin{tabular}{|c|cccccc|c|}
\hline
Level & BSk2 & BSk3 & BSk6 & MSk5 & MSk5$^*$ &SLy4& Expt.\\
\hline
$1s_{1/2}$  & -38.7 & -37.1 & -44.2 &-37.5  &-48.1 &-49.0 & -- \\
\ldots & \ldots & \ldots & \ldots & \ldots & \ldots & \ldots &\ldots\\
$2p_{1/2}$  & -16.0 & -15.9 & -16.4 & -15.9  & -18.7 & -17.6&-16.1  \\
$1g_{9/2}$* & -14.7 & -14.6 & -14.9 & -14.7 & -16.6 & -15.6&-15.8  \\
$1g_{7/2}$  &  -9.5 &  -9.6 &  -9.4 & -9.6  & -11.1  & -9.3&-9.7  \\
$2d_{5/2}$  &  -9.5 &  -9.7 &  -8.9 & -9.7  & -10.9 & -9.2&-8.7 \\
$2d_{3/2}$  &  -7.6 &  -7.9 &  -6.7 & -7.9  & -8.5 & -6.9&-7.2 \\
$3s_{1/2}$  &  -7.3 &  -7.6 &  -6.3 & -7.6   & -8.7  &-6.4&--  \\
$1h_{11/2}$  &  -7.1 &  -7.3 &  -6.2 & -7.4  & -7.5  &-6.2&-6.8  \\
\hline
$\Delta_p$  &3.7  &3.4  &4.5  &3.7  &4.3  & 4.5&5.1 \\
\hline
\end{tabular}
\end{center}
\end{table}

\begin{table}[b!]
 \centering
 \caption{Single-particle neutron levels in $^{132}$Sn (MeV). 
Experimental values are taken from Ref.\ \cite{mez95}. The asterisk denotes
the Fermi level. The quantity $\Delta_n$ is the interval between the centroids 
of the $2d$ and $3p$ doublets. }
\label{tab_sp4}
\begin{center}
\begin{tabular}{|c|cccccc|c|}
\hline
Level & BSk2 & BSk3 & BSk6 & MSk5 & MSk5$^*$ &SLy4& Expt.\\
\hline
$1s_{1/2}$   &  -37.8 & -35.3 & -49.8 &-39.0 &-48.1 & -55.8&--  \\
\ldots & \ldots & \ldots & \ldots & \ldots & \ldots & \ldots\ldots &\\
$1g_{7/2}$   &   -9.2 &  -8.8 & -11.2 &-9.5  &-11.1  &-11.4&-9.8  \\
$2d_{5/2}$   &  -10.2 &  -9.9 & -11.2 &-10.3  &-10.9 &-11.7&-9.0  \\
$3s_{1/2}$   &   -8.3 &  -8.1 &  -8.9 &-8.4   &-8.7 &-9.4&-7.7  \\
$1h_{11/2}$  &   -7.3 &  -7.2 &  -7.5 & -7.1  &-7.5 & -7.7&-7.6 \\
$2d_{3/2}$*  &   -7.9 &  -7.8 &  -8.6 &-8.0   &-8.5 &-9.1&-7.4  \\
$2f_{7/2}$   &   -2.9 &  -3.1 &  -2.3 & -2.8  &-2.2  &-2.0&-2.4  \\
$3p_{3/2}$   &   -1.3 &  -1.5 &  -0.7 & -1.2  &-0.5  &-0.3&-1.6  \\
$1h_{9/2}$   &   0.0 &  -0.1 &   0.2 & -0.1   &0.1 &0.8&-0.9  \\
$3p_{1/2}$   &   -0.7 &  -0.9 &  0.0 & -0.5  &0.1 &0.3&-0.8  \\
$2f_{5/2}$   &   -0.4 &  -0.6 &   0.4 & -0.3   &0.4  &0.6&-0.4  \\
\hline
$\Delta_n$  &8.2  &7.8  &9.7  &8.4  &9.6 & 10.6& 7.1\\
\hline
\end{tabular}
\end{center}
\end{table}

\begin{table}[t!]
 \centering
 \caption{Single-particle proton levels in $^{16}$O (MeV).
Experimental values are taken from Ref.\ \cite{vb72}. The asterisk denotes
the Fermi level. The quantity $\Delta_p$ is the interval between the centroid
of the $1p$ doublet and the $2s_{1/2}$ state. (Errors in Table 6a of Ref.
\cite{far01} for MSk5$^*$ corrected.)}
\label{tab_sp5}
\begin{center}
\begin{tabular}{|c|cccccc|c|}
\hline
Level & BSk2 & BSk3 & BSk6 & MSk5 & MSk5$^*$ & SLy4&Expt.\\
\hline
$1s_{1/2}$  & -25.5 & -24.5 & -29.4 & -25.4 & -30.3 &-32.9& -40 $\pm$8\\
$1p_{3/2}$  & -15.1 & -14.8 & -15.8 & -15.1 & -16.6 & -17.2&-18.4\\
$1p_{1/2}$* &  -9.5 &  -9.3 &  -9.4 &  -9.6 &  -10.6 & -11.1&-12.1\\
$1d_{5/2}$  &  -4.3 &  -4.4 &  -3.1 &  -4.3 &  -3.6 &-3.4& -0.6\\
$2s_{1/2}$  &  -1.7 &  -2.0 &  -0.4 &  -1.7 &  -0.7 &-1.1& -0.1\\
\hline
$\Delta_p$  &11.5  &11.0  &13.2 &11.6&13.9 &14.1& 16.2\\
\hline
\end{tabular}
\end{center}
\end{table}

\begin{table}[t!]
 \centering
 \caption{Single-particle neutron levels in $^{16}$O (MeV).
Experimental values are taken from Ref.\ \cite{vb72}. The asterisk denotes
the Fermi level. The quantity $\Delta_n$ is the interval between the centroid
of the $1p$ doublet and the $2s_{1/2}$ state.}
\label{tab_sp6}
\begin{center}
\begin{tabular}{|c|cccccc|c|}
\hline
Level & BSk2 & BSk3 & BSk6 & MSk5 & MSk5$^*$ & SLy4&Expt.\\
\hline
$1s_{1/2}$  & -28.9 & -27.9 & -33.0 & -28.9 & -33.9 &-36.7& -\\
$1p_{3/2}$  & -18.5 & -18.2 & -19.2 & -18.5 & -20.1 & -20.7&-21.8\\
$1p_{1/2}$* & -12.7 & -12.6 & -12.7 & -12.9 & -13.9 & -14.5&-15.7\\
$1d_{5/2}$  &  -7.5 &  -7.7 &  -6.3 &  -7.5 &  -6.9 & -6.6&-4.1\\
$2s_{1/2}$  &  -4.8 &  -5.1 &  -3.3 &  -4.8 &  -3.7 & -4.0&-3.3\\
\hline
$\Delta_p$  &11.8  &11.2  &13.7 &11.8&14.8 & 14.6&16.5\\
\hline
\end{tabular}
\end{center}
\end{table}

{\it Single-particle spectra.} Clearly, it is of interest to see what happens 
to the s.p.\ energies when the effective mass is reduced while maintaining a 
very good fit to the masses. In Tables \ref{tab_sp1} -- \ref{tab_sp6} 
and Figure \ref{fig_sp} we show the s.p.\ spectra of $^{208}$Pb, $^{132}$Sn, 
and $^{16}$O for forces BSk2, BSk3, and BSk6, along with those of the old 
forces MSk5, MSk5$^*$ and SLy4. Comparing 
BSk2 and BSk3 shows that making the pairing density-dependent has little effect
on the s.p.\ energies. Also, the spectra of BSk2 and MSk5 are very similar, but
different from those of BSk6, MSk5$^*$, and SLy4, which, however,
resemble each other quite closely. That is, the effective mass still determines
the s.p.\ spectra. But while BSk2 and BSk6 have different s.p.\ spectra they 
give very similar fits to the mass data. 
On the other hand, the s.p.\ spectra of BSk6
are very similar to those of MSk5$^*$, but the latter gives a much worse
mass fit. The only way in which we can reconcile this behavior with the
Strutinsky theorem in the form (\ref{8}) is to invoke the $\delta\,n_i$ 
quantities: shifts in the $\delta\,n_i$ compensate the differences between the 
s.p.\ spectra of BSk2 and BSk6, and at the same time account for the
different mass predictions of BSk6 and MSk5$^*$, despite their similar
s.p.\ spectra. This interpretation in terms of the occupation numbers is
strengthened by the fact that the decoupling of the mass fits from the fits
to the s.p.\ spectra is made possible only by adjustment of the pairing cutoff.

As for the agreement with the experimental s.p.\ spectra, we see that a value
of $M_s^*/M$ close to 1.0 is favored by the $^{208}$Pb data, while $^{16}$O
favors a value of 0.8. The $^{132}$Sn data are ambiguous, the neutron spectrum
indicating the higher value of $M_s^*/M$, and the proton spectrum the lower
value. 
Presumably, if we took into account the coupling of s.p.\ excitations and
surface-vibration RPA modes for the forces with $M_s^*/M$ = 0.8, as in
Refs.\ \cite{bk68,bg80}, the calculated s.p.\ spectra of heavy nuclei would
be in better agreement with experiment, but we would then have to refit to
the mass data.\vskip2cm

\begin{figure}[t!]
\centerline{\epsfig{figure=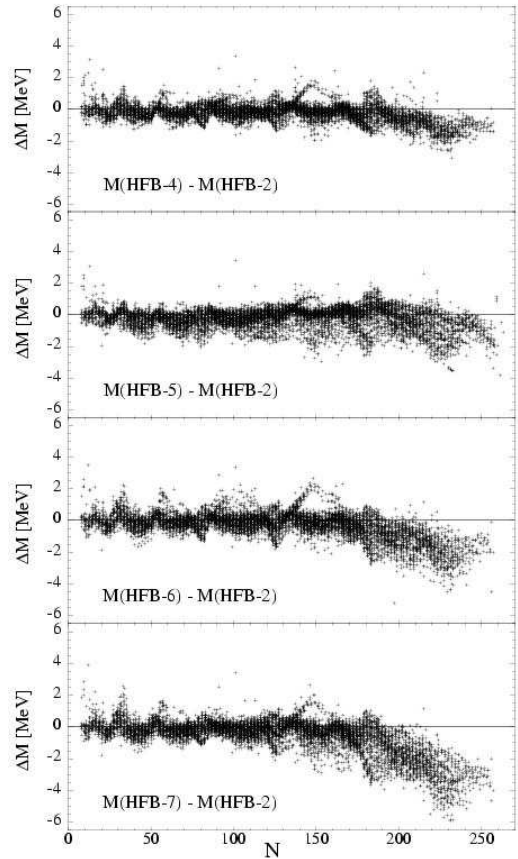,width=8cm}}
\caption{Differences between the HFB-2 and HFB-4 to HFB-7 masses
as a function of the neutron number $N$ for all nuclei with $Z, N \ge$ 8
lying between the proton and neutron driplines up to $Z = 120$.}
\label{fig_mdif}
\end{figure}

\begin{figure}[t!]
\centerline{\epsfig{figure=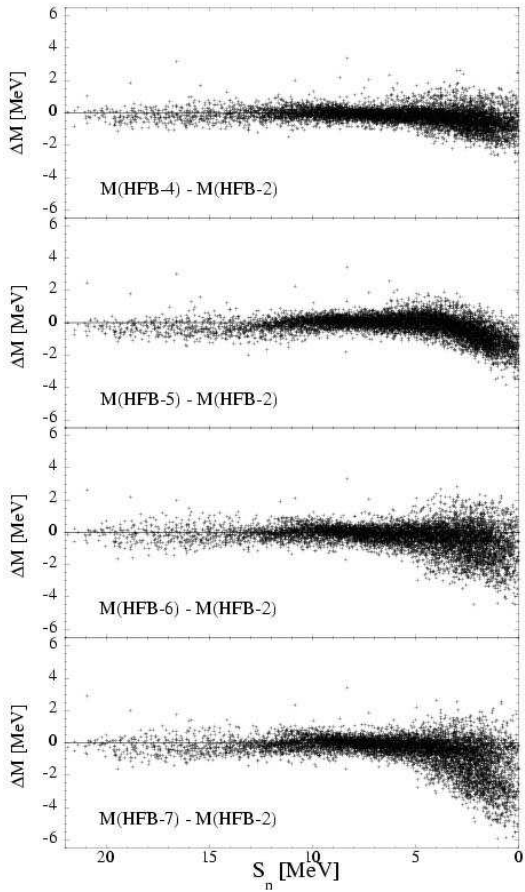,width=8cm}}
\caption{Same as Fig.~\ref{fig_mdif} as a function of the neutron separation energy
$S_n$.}
\label{fig_mdif_sn}
\end{figure}

{\it Extrapolation to drip lines.} With each of the forces BSk4--7 determined 
as described we constructed complete mass tables, labeled HFB-4 to HFB-7, 
respectively, for the same nuclei as were included in the HFB-2 and HFB-3
tables, i.e., all the 9200 nuclei lying between the two drip lines over the 
range of $Z$ and $N \ge 8$ and $Z \le 120$. The differences between the HFB-2 
masses and the HFB-4, 5, 6, and 7 masses are displayed in 
Fig. \ref{fig_mdif} as a function of the neutron number $N$ and in 
Fig. \ref{fig_mdif_sn} as a function of the neutron separation energy $S_n$,  
where it will be seen that these differences never exceed 6 MeV, and 
even then only as the neutron drip line is approached, no matter what the 
value of Z (different published mass formulas all giving very good data fits 
can differ by up to 15
or 20 MeV at the drip lines \cite{rmp03}). Looking at the first
and third panels of  Fig. \ref{fig_mdif} shows that as $M_s^*/M$ is reduced
there is a definite tendency for open-shell nuclei to be bound a little more
strongly, a trend that becomes more conspicuous for the heaviest nuclei. The
second and last panels of this figure confirm the feature already noted in
paper I \cite{ms03} for density-dependent pairing: a similar tendency for
open-shell nuclei to be more strongly bound, especially for the heaviest 
nuclei.

\begin{figure}[t!]
\centerline{\epsfig{figure=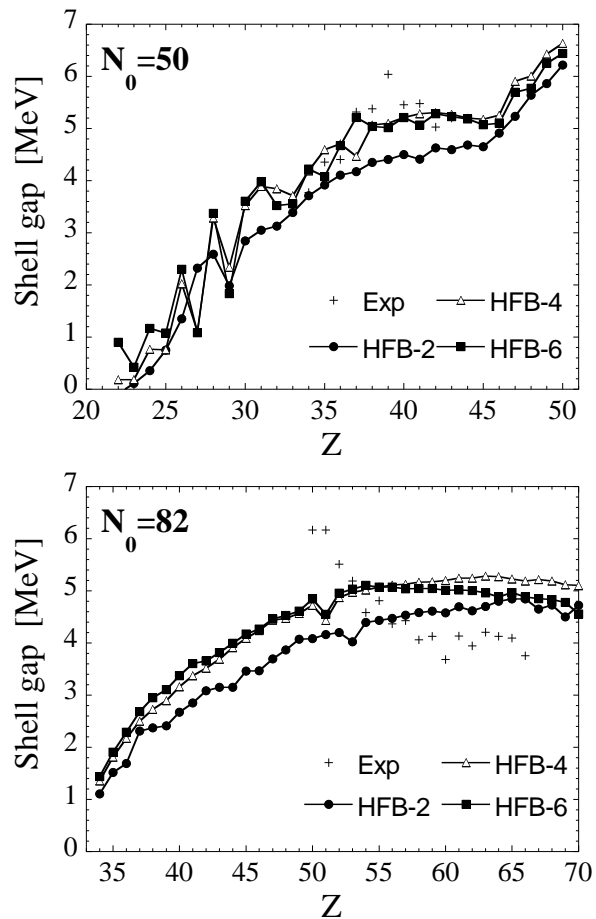,width=8cm}}
\caption{$N_0$ = 50 (upper panel) and $N_0$ = 82 (lower panel) shell gap as a function of
$Z$.}
\label{fig_gap1}
\end{figure}

\begin{figure}[t!]
\centerline{\epsfig{figure=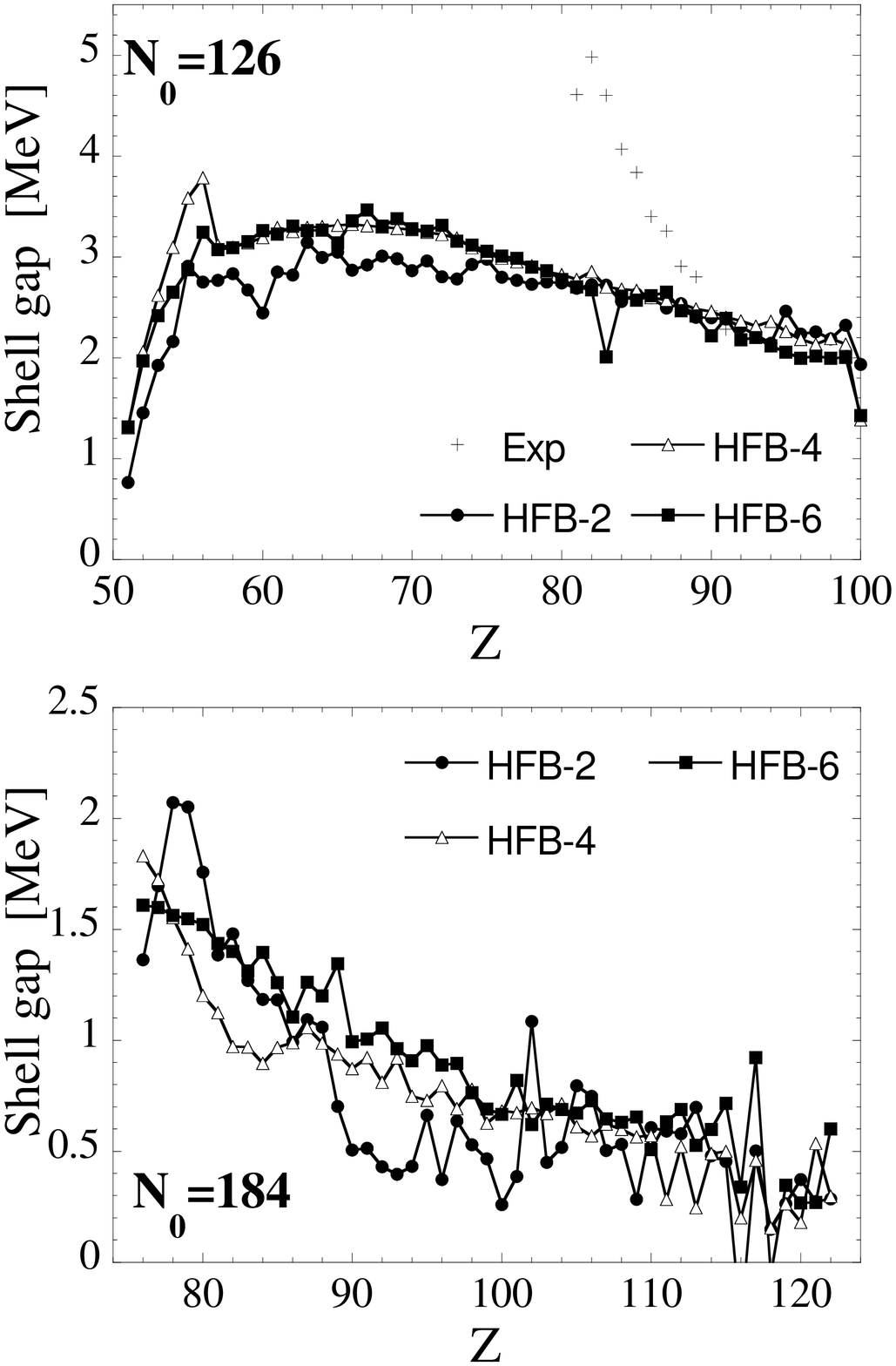,width=8cm}}
\caption{$N_0$ = 126 (upper panel) and $N_0$ = 184 (lower panel) shell gap as a function of
$Z$.}
\label{fig_gap2}
\end{figure}

Actually, while differences of up to 6 MeV between the different mass
predictions for nuclei far from the stability line may appear to be rather 
large, of far greater interest for practical applications such as to the
r-process of nucleosynthesis are differential quantities such as the
neutron separation energy $S_n$.
For these quantities the differences between the different predictions 
are much smaller, as is seen in Figs.\ \ref{fig_gap1} and  \ref{fig_gap2}, 
where we plot as a function of $Z$, for each of the magic numbers
$N_0$ = 50, 82, 126, and 184, respectively, the neutron-shell gaps,
defined by
\begin{eqnarray}
\Delta (N_0) & \equiv & S_{2n}(Z,N_0) - S_{2n}(Z,N_0+2)  \nonumber \\
&=& M(Z,N_0-2) + M(Z,N_0+2) -2M(Z,N_0)\quad  ,
\label{10}
\end{eqnarray}
calculated with BSk2, BSk4 and BSk6 ($S_{2n}$ denotes the 2-neutron separation
energy). From Figs.\ \ref{fig_gap1} and  \ref{fig_gap2}, it can be seen 
that the gaps do not depend significantly on the effective nucleon mass. 
The impact of such
differences on the predicted r-process abundance distributions will 
be studied in more detail in a forthcoming paper.

For $N_0$ = 50, 82, and 126, Figs. \ref{fig_gap1}-\ref{fig_gap2} show a strong
disagreement with experiment for the new mass formulas in the vicinity
of the (semi-) magic {\it proton} numbers $Z$ = 40, 50, and 82. This is related
to the problem of ``mutually supporting magicities" that we have already
discussed in connection with the HFB-2 mass formula \cite{sg02}. Clearly,
it has not been solved, either by reducing the effective mass, or by
introducing the shell-dependent cm correction. (We showed in paper I
\cite{ms03} that making the pairing density-dependent cannot help
in this respect, either.)

\section{Conclusions} 

Fitting Skyrme-type forces to the available mass data without any constraint on
the effective mass always leads to an isoscalar effective mass $M_s^*$ close
to the real nucleon mass $M$. However, we have shown that we can reduce
$M_s^*/M$ to 0.8 without any significant reduction in the quality of the
mass-data fit, although important changes in the s.p.\ spectra are
thereby induced. This decoupling of the fit to the mass data from the fit to
the s.p.\ data was made possible only by exploiting the pairing-force cutoff.

On this basis we constructed four new complete mass tables, referred to as 
HFB-4 to HFB-7, each one including
all the 9200 nuclei lying between the two drip lines over the range of $Z$ and
$N \ge 8$ and $Z \le 120$. HFB-4 and HFB-5 have $M_s^*/M$ constrained to the
value 0.92, with the former having a density-independent pairing, and the 
latter a density-dependent pairing. HFB-6 and HFB-7 are similar, except that
$M_s^*/M$ is constrained to 0.8. The mass-data fits are almost as good as those
given by mass formulas HFB-2 and HFB-3, in which $M_s^*/M$ was unconstrained.  
Actually, in these four new mass formulas we have used an improved treatment
of the center-of-mass correction  \cite{be00}, but although this makes a
difference to individual nuclei we have shown that the overall rms errors
would have been essentially the same if we had used the same correction
as in HFB-2 and HFB-3.

The extrapolations out to the neutron-drip line of all these different mass
formulas are essentially equivalent. We thus see that the mass 
predictions required for the elucidation of the r-process are beginning to
acquire a certain stability against changes in the underlying model.
Nevertheless, it must be remembered that the acquisition of new mass data in
the regions far from stabilty may well necessitate drastic changes to the
underlying model.

Although the forces presented in this paper are equivalent from the standpoint
of nuclear masses, there may still be significant differences as far as
other quantities of astrophysical significance are concerned, e.g., fission
properties, nuclear level densities, giant isovector dipole resonance (GDR), and
beta-strength  functions. Investigations along these lines has already begun, 
and it has
been shown \cite{go03} that the measured positions of the GDR strongly
favor the Skyrme forces BSk6 and BSk7 with their low effective mass of 
$M^*=0.8M$; these calculations were made within the HFB plus 
Quasi-particle Random Phase Approximation (QRPA) framework (the second-RPA 
method being applied to estimate the higher QRPA  effects). However, the 
interpretration 
of such calculations depends on the extra modelling, and the underlying
approximations,  of collective excitations through the QRPA method; 
further studies are needed.

\noindent{\bf Acknowledgements} \\
M.S. and S.G. are FNRS Research Fellow and Associate, respectively. M.B. 
acknowledges support through a European Community Marie Curie fellowship.
We wish to thank P.-H. Heenen for extensive discussions. 
This research was supported in part by the PAI-P5-07 of the Belgian
Office for Scientific Policy and  NSERC (Canada).

\end{document}